\documentclass[12pt,preprint]{aastex}
\usepackage{natbib}
\usepackage{amsmath}
\usepackage{graphicx}
\def\ca{\rm{CA}\,{\small\rmfamily II}\,}
\def\snr{SN\,1993J~}
\def\duesnr{SN\, 2009ig~}
\def\velu{\rm\,{kms^{-1}}}
\def\2F1{~_2F_1}
\def\aap{A\&A\,  }
\def\aj{AJ  }
\def\apj{ApJ\,  }
\def\apjl{ApJ\,  }
\def\apjs{ApJS  }
\def\apss{Astrophysics and Space Science  }

\def\mnras{MNRAS\,  }

\def\nat{Nature\,  }

\def\prd{Phys. Rev. D   }
 
\begin{document}
\shorttitle
{
Relativistic   SN
}
\shortauthors{Zaninetti}
\title
{
A classical and a relativistic law   of
motion for spherical supernovae
}
\author{Lorenzo Zaninetti  }

\affil {Dipartimento di Fisica,
        Via Pietro Giuria 1,\\
        10125 Torino, Italy}

\email   {zaninetti@ph.unito.it  \\
\url     {http://www.ph.unito.it/$\tilde{~}$zaninett}}
\begin{abstract}
In this paper we derive  some  first order differential
equations which model    the classical and the relativistic
thin layer approximations.
The circumstellar medium is
assumed to follow  a density profile of
Plummer type,
or of Lane--Emden ($n=5$) type,
or  a  power law.
The first order differential equations
are solved analytically,
or numerically, or by a series  expansion,
or by recursion.
The initial conditions are chosen in order to model
the temporal evolution of SN 1993J over ten years
and a smaller chi-squared is obtained  for
the  Plummer case with eta=6.
The stellar mass ejected by the SN progenitor 
prior to the explosion,
expressed in solar
mass, is identified with  the total mass associated with
the selected density profile
and varies
from  $0.217 $ to  $0.402 $ when the central number density
is $10^7 $ particles per cubic centimeter.
The  Full width at half maximum of the three density profiles,
which can be identified  with the size of the
Pre-SN 1993J  envelope,
varies
from 0.0071 pc to 0.0092 pc.
\end{abstract}

\section{Introduction}


The absorption features of supernovae (SN)
allow the determination of their expansion velocity,
$v$.
We select, among others, some results.
The  spectropolarimetry
(\ca IR triplet) of SN 2001el
gives a maximum velocity of $\approx 26000\velu$,
see  \cite{Wang2003}.
The same  triplet when searched in seven SN
of type Ia gives
$10400 \velu  \leq v \leq 17700 \velu $,
see Table I in \cite{Mazzali2005}.
A time series of eight  spectra in \duesnr
allows asserting that the velocity at the  \ca line,
for example,
decreases  in  12 days from 32000 $\velu$ to 21500$\velu$,
see Figure 9 in \cite{Marion2013}.
A recent analysis of 58 type Ia SN in Si II
gives  $9660 \velu  \leq v \leq 14820 \velu $,
see Table II
in \cite{Childress2014}.
The  previous analysis  allow saying that
the maximum velocity sofar observed for SN
is  $\frac{v}{c} \approx  0.1$, where
$c$ is the speed of light; this  observational fact
points to a relativistic equation of motion.

We now briefly review the shocks and the
Kompaneyets  approximation in special relativity (SR).
A similar solution for  strong relativistic shocks
in a  circumstellar medium (CSM) which varies with the radius
was found by \cite{Blandford1976}.
Relativistic shocks are commonly used
for gamma ray bursts (GRB) in order to explain
the production of non-thermal electrons,
see  \cite{Baring2011}.
The interactions between the shock and the ambient density
fluctuations can produce turbulence  with
a significant component of magnetic energy,
see  \cite{Inoue2011}.
Relativistic radiation-mediated shocks
can produce GRB with typical  parameters similar
to those observed, see  \cite{Nakar2012}.
Trans-relativistic shocks have
been used to produce high-energy neutrino and gamma-ray in SN,
see \cite{Kashiyama2013}.
The Kompaneyets  approximation is usually
developed in a Newtonian framework,
see \cite{Kompaneets1960,Olano2009},
and has been derived in SR
by  \cite{Shapiro1979,Lyutikov2012}.
The temporal observations of SN such as  \snr
establish a clear   relation
between the instantaneous radius of expansion $r$
and  the time $t$, of the
type  $r \propto t^{0.82}$,
see \cite{Marcaide2009}
and therefore  allows exploring variants of the thin layer
approximation.
The  previous observational facts
excludes  a SN propagation in an CSM
with constant density:
two  solutions of this type are
the  Sedov solution which
scales  as  $r \propto t^{0.4}$,
 see \cite{Sedov1959,Dalgarno1987},
and the  momentum conservation  in  a thin layer
approximation, which
scales  as  $r \propto t^{0.25}$,
see \cite{Dyson1997,Padmanabhan_II_2001}.
Previous efforts  to  model
these observations
in the  framework
of the thin layer approximation
in an CSM governed by a power law,
 see  \cite{Zaninetti2011a},
or in the framework
in which the CSM   has a
constant density but swept mass regulated by a parameter called
porosity,
see \cite{Zaninetti2012c},
have been successfully explored.
An important feature of the various models
is based on the type of CSM  which surrounds the expansion.
As an example in the framework of classical shocks,
\cite{Chevalier1982a} and \cite{Chevalier1982b}
analyzed  self-similar solutions
with an CSM  of the type $r^{-s}$, which means an
inverse power law dependence.
In the framework of the
Kompaneyets
equation, see \cite{Kompaneets1960},
for the motion of a shock wave in different plane-parallel
stratified media,
\cite{Olano2009}  considered four types of CSM.
It is therefore interesting to take into account
an self-gravitating CSM, which gives a physical basis to the
considered model.
The relativistic treatment  has been concentrated on the
determination of the Lorentz factor, $\gamma$,
for the ejecta in GRB; we report some  research in this regard:
\cite{Granot2006}  found  $30< \gamma <50$  for a
significant number  of GRBs,
\cite{Peer2007}  found $\gamma=305$ for GRB 970828  and
$\gamma=384$ for  GRB 990510,
\cite{Zou2010} found high values
for the sample of the GRBs considered
$30.5  \gamma < 900$,
\cite{Aoi2010}
in the framework of a high-energy spectral
cutoff originating from the
creation of electron--positron pairs
found $\gamma \approx$ 600 for GRB 080916C,
\cite{Muccino2013}  found  $\gamma \approx 6.7 \times 10^2$
for GRB 090510.
The last phase
of stellar evolution
predicts the production of $^{56}$Ni, see
 \cite {Truran1967,Bodansky1968,Matz1990,Truran2012}
and therefore this type of decay has been used
to model
the light curve of supernovae (SN),
see  among others \cite{Mazzali1997,Elmhamdi2003,
Stritzinger2006,Magkotsios2010,
Krisciunas2011,Okita2012,Chen2013}
as well the
reddening measurements of the supernova remnant (SNR)
Cassiopeia A, see \cite{Eriksen2009}.
These theoretical and observational efforts give interest
to the exploration of the modification of  $^{56}$Ni decay
due to time dilation.


In this paper we
review the  standard
two-phase model for the expansion of a SN,
see Section \ref{standard},
and  three density profiles,
see Section \ref{secdensity}.
In  Section \ref{secclassic}
we derive the  differential equations
which model the thin layer approximation
for a   SN in the presence of three types of medium.
Section  \ref{secclassic}  also contains a model in which
the center of the
explosion does not coincide with the center of the polytrope.
A  relativistic treatment is carried out in
Section \ref{secrelativistic}.
The  application of the developed theory to \snr
is  split into the classical case,
see Section \ref{applicationsclassic},
and the relativistic case,
see Section  \ref{applications}.

\section{The standard model}
\label{standard}

\begin{figure*}
\begin{center}
\includegraphics[width=7cm ]{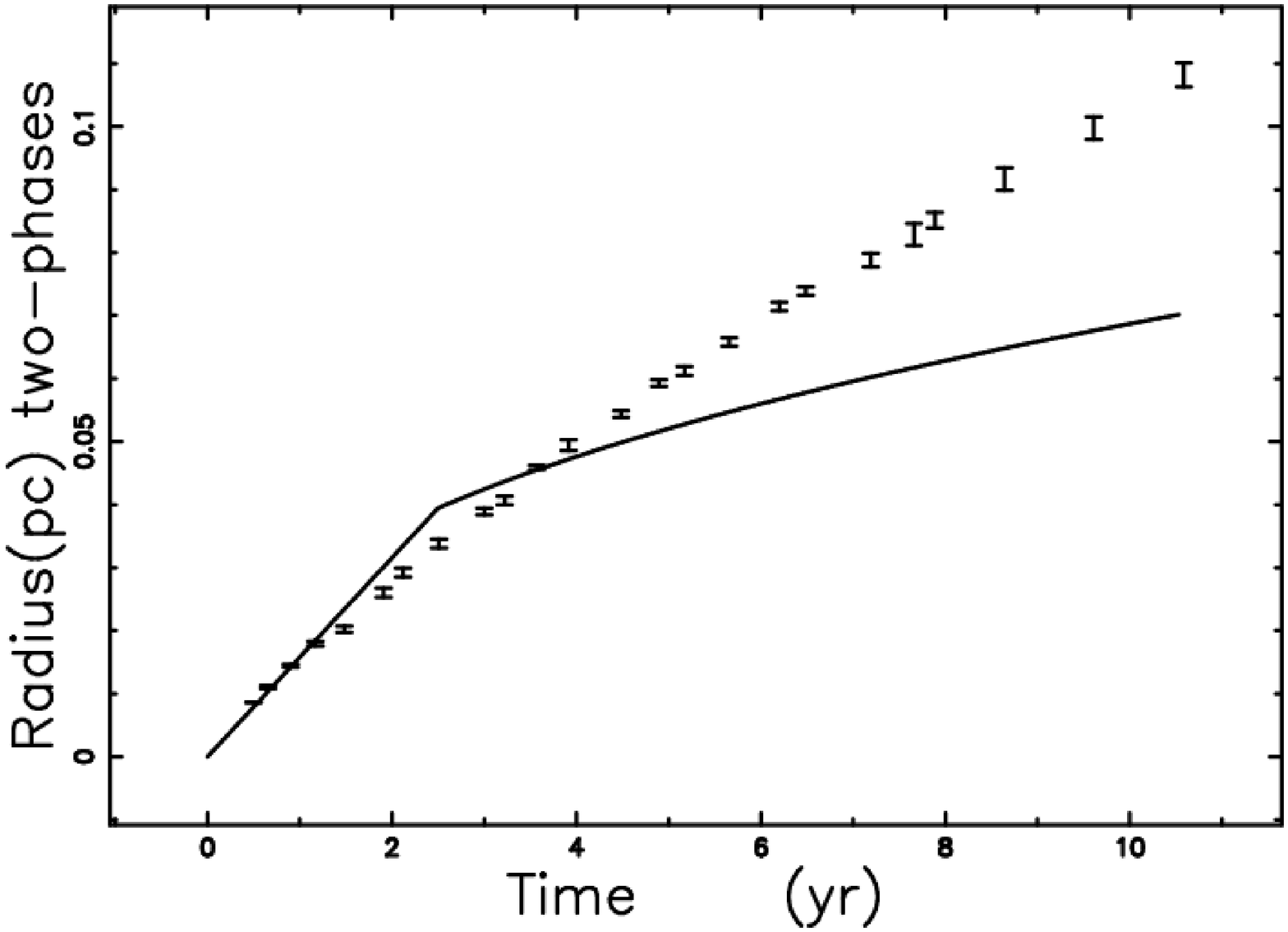}
\end {center}
\caption
{
Theoretical radius as given by the two-phase
solution (full line)
and astronomical data of \snr with
vertical error bars.
}
\label{1993duefasi}
    \end{figure*}

A SN  expands at a constant velocity until
the surrounding mass is
of the order of the solar mass.
This time, $t_M$,
is
\begin {equation}
t_M= 186.45\,{\frac {\sqrt [3]{{\it M_{\sun} }}}{\sqrt [3]{{\it n_0}}{\it
v_{10000}}}} \quad yr
\quad ,
\end{equation}
where $M_{\sun}$ is the number of solar masses
in the volume occupied by the SN,
$n_0$ is  the
number density  expressed  in particles~$\mathrm{cm}^{-3}$,
and $v_{10000}$ is the initial velocity expressed
in units of 10000\ km/s, see \cite{Dalgarno1987}.
A first law of motion for the $SN$
is the  Sedov  solution
\begin{equation}
R(t)=
 ({\frac {25}{4}}\,{\frac {{\it E}\,{t}^{2}}{\pi \,\rho}}  )^{1/5}
\quad ,
\label{sedov}
\end{equation}
where $E$ is the energy injected into the process
and $t$ is the  time,
see~\cite{Sedov1959,Dalgarno1987}.
Our astrophysical  units are: time, ($t_1$), which
is expressed  in years;
$E_{51}$, the  energy in  $10^{51}$ erg;
$n_0$,  the
number density  expressed  in particles~$\mathrm{cm}^{-3}$~
(density~$\rho=n_0m$, where $m = 1.4m_{\mathrm {H}}$).
In these units, Eq.~(\ref{sedov}) becomes
\begin{equation}
R(t) \approx  0.313\,\sqrt [5]{{\frac {{\it E_{51}}\,{{\it t_1}}^{2}}{{\it n_0}}}
}~{pc}
\quad .
\label{sedovastro}
\end{equation}
The Sedov solution scales as $t^{0.4}$.
We are now ready to couple
the Sedov phase with the free expansion phase
\begin{equation}
 R(t)  = \{ \begin{array}{ll}
0.0157t\,\mbox{pc} &
\mbox {if $t \leq 2.5$ yr } \\
0.0273\,\sqrt [5]t^2\, \mbox{pc}   &
\mbox {if $t >    2.5$ yr.}
            \end{array}
\label{twophases}
\nonumber
\end{equation}
This two-phase solution is obtained with
the following parameters
$M_{\sun}$ =1 ,
$n_0 =1.127 \times10^5$,
$E_{51}=0.567$
and Fig. \ref{1993duefasi} presents its  temporal behavior
as well as the data.
A similar  model  is reported in \cite{Spitzer1978} with the
difference that the first phase ends  at $t=60$\ yr against our
$t=2.5$\ yr.
A careful  analysis of Fig. \ref{1993duefasi}   reveals that
the standard two-phase model does not fit
the observed radius--time relation  for  \snr.

\section{Density profiles for the CSM}

\label{secdensity}

This  section  introduces three density profiles
for the CSM:
a Plummer-like profile,
an self-gravitating profile  of Lane--Emden type,
and   a power law profile .

\label{secmomentum}

\subsection{The Plummer profile}
The Plummer-like  density profile, after \cite{Plummer1911},
is
\begin{equation}
\rho(r;R_{flat}) =
\rho_c  ({\frac {R_{flat}}{({R_{flat}^2 +r^2})^{1/2}}}  )^{\eta}
\label{densitaplummerflat}
\nonumber
\quad
\end{equation}
where $r$ is the distance from the center,
$\rho$    is the density,
$\rho_c$  is the density at the center,
$R_{flat}$ is the distance before  which the density is nearly constant,
and $\eta$  is the power law exponent at large values of $r$,
see \cite{Whitworth2001} for more details.
The following transformation, $R_{flat}= \sqrt{3} b $,
gives the  Plummer-like profile, which can be compared
with the Lane--Emden profile
\begin{equation}
\rho(r;b) =
\rho_c \bigl ( \frac{1}{1+ \frac{1}{3}\,{\frac {{r}^{2}}{{b}^{2}}}}
 \bigr ) ^{\eta/2}
\quad .
\label{densitaplummer}
\quad
\end{equation}
At low values of $r$,  the Taylor expansion of the Plummer-like profile
can be taken:
 \begin{equation}
\rho(r;b) \approx \rho_{{c}}( 1-1/6\,{\frac {\eta\,{r}^{2}}{{b}^{2}}})
\label{scalingplummerlow}
\nonumber
\quad ,
\end{equation}
and at  high  values of $r$,  the
behavior  of the Plummer-like profile is
\begin{equation}
\rho(r;b) \sim \rho_c
( \sqrt{3}\,b ) ^{\eta} ( \frac{1}{r} ) ^{
\eta}
\nonumber
\label{scalingplummerhigh}
\quad .
\end{equation}

The  total  mass $M(r;b)$ comprised between
0 and  $r$
is
\begin{equation}
M(r;b) = \int_0^r   4 \pi r^2   \rho(r;b) dr
= \frac{PN}{PD}
\quad ,
\end{equation}
where
\begin{eqnarray}
PN =-\sqrt {3}b\rho_{{c}}\pi \, ( 4\,
{\mbox{$_2$F$_1$}(\eta/2,-3/2+\eta/2;\,-1/2+\eta/2;\,-3\,{\frac {{b}^{2}}{{r}^{2}}})} \times  \nonumber \\
\times \Gamma  ( -\eta/2+5/2 ) \Gamma  ( \eta/2 ) \cos
 ( 1/2\,\pi \,\eta ) {r}^{3-\eta}{3}^{1/2+\eta/2}{b}^{\eta-
1} \nonumber \\
-9\,{\pi }^{3/2}{b}^{2}\eta+27\,{\pi }^{3/2}{b}^{2} )
\nonumber
\end{eqnarray}
and
\begin{equation}
PD=3\,\cos ( 1/2\,\pi \,\eta ) \Gamma  ( -\eta/2+5/2
 )  ( \eta-3 ) \Gamma  ( \eta/2 )
\nonumber
\quad ,
\end{equation}
where ${\2F1(a,b;\,c;\,z)}$ is the
regularized hypergeometric
function
\cite{Abramowitz1965,NIST2010}.
The above expression simplifies when $\eta=6$ , $M(r;b)_6$,
\begin{eqnarray}
M(r;b)_6 =
{\frac {27\,\rho_{{c}}\pi \,{b}^{7}\sqrt {3}}{2\, ( 3\,{b}^{2}+{r
}^{2} ) ^{2}}\arctan ( 1/3\,{\frac {r\sqrt {3}}{b}}
 ) }+9\,{\frac {\rho_{{c}}\pi \,{b}^{5}\sqrt {3}{r}^{2}}{
 ( 3\,{b}^{2}+{r}^{2} ) ^{2}}\arctan ( 1/3\,{\frac {r
\sqrt {3}}{b}} ) }+ \nonumber \\
3/2\,{\frac {\rho_{{c}}\pi \,{b}^{3}\sqrt {3}
{r}^{4}}{ ( 3\,{b}^{2}+{r}^{2} ) ^{2}}\arctan ( 1/3\,{
\frac {r\sqrt {3}}{b}} ) }-{\frac {27\,\rho_{{c}}\pi \,{b}^{6}r
}{2\, ( 3\,{b}^{2}+{r}^{2} ) ^{2}}}+9/2\,{\frac {\rho_{{c}}
\pi \,{b}^{4}{r}^{3}}{ ( 3\,{b}^{2}+{r}^{2} ) ^{2}}}
 \quad .
 \nonumber
\end{eqnarray}
The astrophysical version of the total mass is
\begin{equation}
M(r_{pc};b_{pc}) = \frac{PNA}{PDA}  \,M_{\sun}
\quad ,
\nonumber
\end{equation}
with
\begin{eqnarray}
PNA= - 2.47\,10^{-10}\,{b_{{{\it pc}}}}^{3}n_{{0}} [
 1.02\,10^{10}\,\arctan (  1.73\,{\frac {b_{{{\it pc}}}}{r
_{{{\it pc}}}}}  ) {b_{{{\it pc}}}}^{4}
 \nonumber \\
+ 6.8\,10^9\,\arctan
 (  1.73\,{\frac {b_{{{\it pc}}}}{r_{{{\it pc}}}}}
  ) {b_{{{\it pc}}}}^{2}{r_{{{\it pc}}}}^{2}
  + 1.13\,10^9\,
\arctan (  1.73\,{\frac {b_{{{\it pc}}}}{r_{{{\it pc}}}}}
  ) {r_{{{\it pc}}}}^{4}- 1.6\,10^{10}\,{b_{{{\it pc}}}}^{4}
\nonumber \\
  +  5.89\,10^9\,{b_{{{\it pc}}}}^{3}r_{{{\it pc}}}- 1.06\,10^{10}\,{b_{
{{\it pc}}}}^{2}{r_{{{\it pc}}}}^{2}- 1.96\, 10^9\,b_{{{\it pc}}}{r_{
{{\it pc}}}}^{3}- 1.78\,10^9\,{r_{{{\it pc}}}}^{4}  ]
\nonumber
\end{eqnarray}
and
\begin{equation}
PDA=\left(  3.0\,{b_{{{\it pc}}}}^{2}+{r_{{{\it pc}}}}^{2} \right) ^{2}
\quad ,
\nonumber
\end{equation}
where ${b_{{{\it pc}}}}$ is b expressed in pc,
${r_{{{\it pc}}}}$ is r expressed in pc
and $n_0$ is the same of
eqn.(\ref{sedovastro}).
The relationship between Full width at half maximum (FWHM)  and  ${b_{{{\it pc}}}}$
 is
\begin{equation}
 FWHM=1.766\,b_{{{\it pc}}}
\quad .
\nonumber
\end{equation}

\subsection{The Lane--Emden profile}

The self gravitating sphere of polytropic
gas is governed
by the Lane--Emden differential equation
of the second order
\begin{equation}
{\frac {d^{2}}{d{x}^{2}}}Y ( x ) +2\,{\frac {{\frac {d}{dx}
}Y ( x ) }{x}}+ ( Y ( x )  ) ^{n}=0
\quad ,
\nonumber
\end{equation}
where $n$  is an integer, see
\cite{Lane1870,Emden1907,Chandrasekhar_1967,Binney2011,Zwillinger1989}.

The solution  $Y ( x )_n$
has  the density profile
\begin{equation}
\rho = \rho_c Y ( x )_n^n
\quad ,
\nonumber
\end{equation}
where $\rho_c$ is the density at $x=0$.
The pressure $P$ and temperature $T$ scale as
\begin{equation}
P = K \rho^{1 +\frac{1}{n}}
\quad ,
\label{pressure}
\end{equation}
\begin{equation}
 T = K^{\prime} Y(x)
\label{temperature}
\quad ,
\end{equation}
where  K and K$^{\prime}$   are two  constants,
for more details, see \cite{Hansen1994}.

Analytical solutions exist for $n=0$, 1 and 5;
that  for $n$=0 is
\begin{equation}
Y(x) = \frac{sin(x)}{x}
\quad,
\nonumber
\end{equation}
and has therefore an oscillatory behavior.
The analytical  solution for $n$=5 is
\begin{equation}
Y(x) ={\frac {1}{{(1+ \frac{{x}^{2}}{3})^{1/2}}} }
\quad ,
\nonumber
\end{equation}
and the density for $n$=5 is
\begin{equation}
\rho(x) =\rho_c {\frac {1}{{(1+ \frac{{x}^{2}}{3})^{5/2}}} }
\label{densita5}
\quad .
\end{equation}

The   variable  $x$   is   non-dimensional
and  we now  introduce the
new variable $x=r/b$
\begin{equation}
\rho(r;b) =\rho_c {\frac {1}{{(1+ \frac{{r}^{2}}{3b^2})^{5/2}}} }
\label{densita5b}
\nonumber
\quad .
\end{equation}
This profile is a particular case, $\eta=5$,
of the Plummer-like profile as given by Eq. (\ref{densitaplummer}).
At low values of $r$,  the Taylor expansion of this profile is
 \begin{equation}
\rho(r;b) \approx \rho_{{c}} ( 1-5/6\,{\frac {{r}^{2}}{{b}^{2}}} )
\nonumber
\label{scalinglow}
\quad ,
\end{equation}
and at  high  values of $r$,  its
 behavior is
\begin{equation}
\rho(r;b) \sim 9\,{\frac {\rho_{{c}}\sqrt {3}{b}^{5}}{{r}^{5}}}
\label{scalinghigh}
\quad .
\end{equation}
The FWHM is
\begin{equation}
 FWHM=1.95\,b_{{{\it pc}}}
\quad .
\nonumber
\end{equation}
The gradient here is assumed to be local: it covers lengths smaller than 1 pc,
and is not connected with the gradient which regulates the equilibrium of a
galaxy that extends over a region of some kpc.
The interaction of the progenitor star
with the CSM, through stellar winds, creates a CSM. The
dynamics of this medium can be far equilibrium. The main
astrophysical assumption adopted here is that the density of the CSM
decreases smoothly at low values of distance, as $\rho \approx A -B\,r^2$, due to
previous stellar winds, and $rho \approx C\,r^{-5}$ in the far regions not contaminated
by the previous activity; $A$, $B$, and $C$ being constants.
In view of the behavior of this self-gravitating  profile
at high $r$,   Section \ref{powerlawsubsec} will
analyze a power law dependence for the CSM.

The  total mass $M(r;b)$ comprised between
0 and  $r$
is
\begin{equation}
M(r;b) = \int_0^r   4 \pi r^2   \rho(r;b) dr
=\frac
{
4\,{b}^{3}{r}^{3}\rho_c\,\pi \,\sqrt {3}
}
{
( 3\,{b}^{2}+{r}^{2} ) ^{3/2}
}
\quad ,
\label{integralemassa}
\end{equation}
or  in solar units
\begin{equation}
M(r_{pc};b_{pc}) =
\frac
{
{ 2.2\times 10^{55}}\,{b_{{{\it pc}}}}^{3}{r_{{{\it pc}}}}^{3}
n_{{0}}
}
{
\left( { 2.85\times 10^{37}}\,{b_{{{\it pc}}}}^{2}+{
 9.52\times 10^{36}}\,{r_{{{\it pc}}}}^{2} \right) ^{3/2}
}
 \,M_{\sun}
\quad .
\nonumber
\end{equation}
The total mass of the profile  can be found
calculating the limit for $r \to \infty$
of equation (\ref{integralemassa})
\begin{equation}
M(\infty;b) = \lim_{r\to \infty}M(r;b) = 4\,{b}^{3}{\it \rho_c}\,\pi \,\sqrt
{3}
\label{totalmass}
\end{equation}
Another interesting  physical quantity
deduced in the framework of the virial theorem
is the mean
square speed of the system,
which according
to formula (4.249a) in \cite{Binney2011}  is
\begin{equation}
<v^2> = \frac{GM}{r_g}
\quad ,
\label{vquadromedio}
\end{equation}
where $M$ is the total mass,
$r_g$  is the gravitational radius as defined in
equation (2.42) in \cite{Binney2011}
and G  is the Newtonian gravitational constant.
In the case of a
Lane--Emden profile  as given by equation (\ref{densita5b})
the gravitational radius is
\begin{equation}
r_g = \frac{32\,b\sqrt {3}}
           {3\,\pi }
\label{gravitationalradius}
\quad .
\end{equation}
The mean
square speed of the system
according to formulae
(\ref{gravitationalradius}) and (\ref{totalmass})
is
\begin{equation}
<v^2> = \frac{
3\,G{b}^{2}\rho_{{c}}{\pi }^{2}
}
{
8
}
\label{vsquaremean}
\quad .
\end{equation}
The relationship between  gravitational radius,
$r_g$,  and half mass radius, $r_h$,
is
\begin{equation}
\frac{r_h}{r_g}= 0.16647
\quad ,
\end{equation}
and this allows a more simple  definition for the
mean  square speed
\begin{equation}
<v^2> = 0.16647\,\frac{GM}{r_h}
\quad .
\label{vquadromediorh}
\end{equation}
The astrophysical version of the square root of the
mean square speed as given by formula (\ref{vsquaremean})
is
\begin{equation}
\sqrt{<v^2>} =
233.81 \,\sqrt {{b_{{{\it pc}}}}^{2}n_{{7}}} \,\, \frac{km}{s}
\quad ,
\label{vvirial}
\end{equation}
where  $b_{pc}$ is the scale  parameter expressed in pc,
$n_7$ represents the number density expressed
in $10^7 \mathrm{cm}^{-3}$ units
and  $G=6.67384\, m^3 kg^{-1} s^{-2}$,
see  \cite{CODATA2012}.
\subsection{A power law for the CSM}
\label{powerlawsubsec}

We now  assume that the CSM
around the SN
scales with the following piecewise dependence
(which avoids a pole at $r=0$)
\begin{equation}
 \rho (r;r_0,d)  = \{ \begin{array}{ll}
            \rho_c                      & \mbox {if $r \leq r_0 $ } \\
            \rho_c (\frac{r_0}{r})^d    & \mbox {if $r >    r_0 $.}
            \end{array}
\label{piecewise}
\end{equation}

The mass swept, $M_0$,
in the interval [0,$r_0$]
is
\begin{equation}
M_0 =
\frac{4}{3}\,\rho_{{0}}\pi \,{r_{{0}}}^{3}
\quad .
\nonumber
\end{equation}
The total mass swept, $ M(r;r_0,d) $,
in the interval [0,r]
is
\begin{eqnarray}
M (r;r_0,d)=
-4\,{r}^{3}\rho_{{c}}\pi \, ( {\frac {r_{{0}}}{r}} ) ^{d}
 (d -3 ) ^{-1}   \nonumber \\
+4\,{\frac {\rho_{{c}}\pi \,{r_{{0}}}^{3}}{d-
3}}
+ \frac{4}{3}\,\rho_{{c}}\pi \,{r_{{0}}}^{3}
\quad .
\nonumber
\label{masspowerlaw}
\end{eqnarray}
or  in solar units
\begin{equation}
M(r_{pc};{r_{{0,{\it pc}}}},d) =
\frac
{
 3.14\,n_{{0}} \left(  0.137\,{r_{{{\it pc}}}}^{3}
 \left( {\frac {r_{{0,{\it pc}}}}{r_{{{\it pc}}}}} \right) ^{d}-
 0.0459\,{r_{{0,{\it pc}}}}^{3}d \right)
}
{
3-d
}
\,M_{\sun}
\quad ,
\nonumber
\end{equation}
where  ${r_{{0,{\it pc}}}}$ is  $r_0$ expressed in pc.
The FWHM  is
\begin{equation}
 FWHM=
 \frac
 {
 2\,r_{{0,{\it pc}}}
 }
 {
 {{\rm e}^{-{\frac {\ln  \left( 2 \right) }{d-2}}}}
 }
 \quad .
 \nonumber
 \end{equation}

\section{Classical conservation of momentum}

\label{secclassic}

This section reviews  the standard
equation of motion in the case of
the thin layer approximation in the presence of an CSM with constant density
and derives the equation of motion  under the conditions
of each of the three density profiles for the  density of the CSM.
A simple asymmetrical model
is  introduced.

\subsection{Motion with constant density}

In the case of a constant  density of the CSM,
$\rho_c$, the differential equation
which models momentum conservation
is
\begin{equation}
\frac{4}{3}\,\pi \, ( r ( t )  ) ^{3}\rho_{{c}}{\frac {d
}{dt}}r ( t ) -\frac{4}{3}\,\pi \,{{\it r_0}}^{3}\rho_{{c}}v_{{0}}=0
\quad ,
\nonumber
\end{equation}
where the initial  conditions
are  $r=r_0$  and   $v=v_0$
when $t=t_0$.
The variables can be separated and
the radius as a function of the time
is
\begin{equation}
r(t)= \sqrt [4]{4\,{r_{{0}}}^{3}v_{{0}} ( t-t_{{0}} ) +{r_{{0}}}^
{4}}
\quad ,
\nonumber
\end{equation}
and its   behavior
as $\quad t  \rightarrow \infty$  is
\begin{eqnarray}
r(t) =\sqrt {2}{r_{{0}}}^{3/4}\sqrt [4]{v_{{0}}}\sqrt [4]{t-t_{{0}}}+ \frac{1}{16}\,{
\frac {\sqrt {2}{r_{{0}}}^{7/4}}{{v_{{0}}}^{3/4} ( t-t_{{0}}
 ) ^{3/4}}}.
\nonumber
\end{eqnarray}
The velocity  as a function of time is
\begin{equation}
v(t)=
\frac
{
{r_{{0}}}^{3}v_{{0}}
}
{
 ( 4\,{r_{{0}}}^{3}v_{{0}} ( t-t_{{0}} ) +{r_{{0}}}^{4
} ) ^{3/4}
}
\quad .
\nonumber
\end{equation}

\subsection{Motion with Plummer profile}

The case of a Plummer-like profile  for  the  CSM
as given by  (\ref{densitaplummer})
when   $\eta=6$
produces the differential equation
\begin{equation}
\frac{d}{dt}r(t) = \frac {NDEP}{DNEP}
\quad ,
\label{eqndiffplummer}
\end{equation}
where
\begin{eqnarray}
NDEP =
 ( 9\,\sqrt {3}\arctan  ( 1/3\,{\frac {{\it r0}\,\sqrt {3}}{b
}}  ) {b}^{4}+6\,\sqrt {3}\arctan  ( 1/3\,{\frac {{\it r0}\,
\sqrt {3}}{b}}  ) {b}^{2}{{\it r0}}^{2} +
\nonumber \\
+\sqrt {3}\arctan  ( 1
/3\,{\frac {{\it r0}\,\sqrt {3}}{b}}  ) {{\it r0}}^{4}-9\,{b}^{3}
{\it r0}+3\,b{{\it r0}}^{3}  ) {\it v0}\,  ( 3\,{b}^{2}+
  ( r  ( t  )   ) ^{2}  ) ^{2}
\quad ,
\nonumber
\end{eqnarray}
and
\begin{eqnarray}
NDEP =
 ( 3\,{b}^{2}+{{\it r0}}^{2}  ) ^{2}  ( 9\,\sqrt {3}
\arctan  ( 1/3\,{\frac {r  ( t  ) \sqrt {3}}{b}}  )
{b}^{4}+6\,\sqrt {3}\arctan  ( 1/3\,{\frac {r  ( t  )
\sqrt {3}}{b}}  ) {b}^{2}  ( r  ( t  )   ) ^{2}
\nonumber \\
+\sqrt {3}\arctan  ( 1/3\,{\frac {r  ( t  ) \sqrt {3}}{b}
}  )   ( r  ( t  )   ) ^{4}-9\,{b}^{3}r  (
t  ) +3\,b  ( r  ( t  )   ) ^{3}  )
\quad .
\nonumber
\end{eqnarray}
There is no analytical solution to this differential equation,
but the  solution can be found numerically.

\subsection{Motion with Lane--Emden profile}

In the case of variable  density for  the  CSM
as given by  the profile (\ref{densita5}),
the differential equation
which models momentum conservation is
\begin{equation}
4\,{\frac {{b}^{3} ( r ( t )  ) ^{3}\rho_{{c}}
\pi \,\sqrt {3}{\frac {d}{dt}}r ( t ) }{ ( 3\,{b}^{2}+
 ( r ( t )  ) ^{2} ) ^{3/2}}}-4\,{\frac {{
b}^{3}{r_{{0}}}^{3}\rho_{{c}}\pi \,\sqrt {3}v_{{0}}}{ ( 3\,{b}^{2
}+{r_{{0}}}^{2} ) ^{3/2}}}=0
\quad .
\label{eqndiff}
\end{equation}
The variables can be separated and the solution  is
\begin{equation}
r(t;r_0,v_0,t_0,b) =
\frac{N}{D}
\quad  ,
\label{radiusemdent}
\end{equation}
where
\begin{eqnarray}
N  =
\sqrt {2}{r_{{0}}}^{3/4}\bigl ({r_{{0}}}^{13/2}+2\,{r_{{0}}}^{11/2}
 ( t-t_{{0}} ) v_{{0}}
\nonumber \\
+{r_{{0}}}^{9/2} ( t-t_{{0}}
 ) ^{2}{v_{{0}}}^{2}+6\,{b}^{2}{r_{{0}}}^{9/2}
\nonumber \\
+18\,{b}^{2}{r_{{0
}}}^{7/2} ( t-t_{{0}} ) v_{{0}}+\sqrt {A}{r_{{0}}}^{4}+
\sqrt {A}{r_{{0}}}^{3} ( t-t_{{0}} ) v_{{0}}
\nonumber   \\
+9\,{b}^{4}{r_{
{0}}}^{5/2}+36\,{b}^{4}{r_{{0}}}^{3/2}v_{{0}} ( t-t_{{0}}
 )
\nonumber \\
+9\,\sqrt {A}{b}^{2}{r_{{0}}}^{2}+18\,\sqrt {A}{b}^{4} \bigr )^{1/2}
\nonumber
\end{eqnarray}
and
\begin{eqnarray}
D=2\, ( 3\,{b}^{2}+{r_{{0}}}^{2} ) ^{3/2}
\nonumber
\quad  ,
\end{eqnarray}
with
\begin{eqnarray}
A(t-t_0)=
{r_{{0}}}^{3} ( t-t_{{0}} ) ^{2}{v_{{0}}}^{2}+36\,{b}^{4}
 ( t-t_{{0}} ) v_{{0}}
\nonumber  \\
+18\,{b}^{2}{r_{{0}}}^{2} ( t-t_
{{0}} ) v_{{0}}
\nonumber \\
+2\,{r_{{0}}}^{4} ( t-t_{{0}} ) v_{{0}
}+9\,{b}^{4}r_{{0}}+6\,{b}^{2}{r_{{0}}}^{3}+{r_{{0}}}^{5}
\nonumber
\quad  .
\end{eqnarray}
This is the {\it first} solution and  has an analytical form.
The  analytical  solution  for the velocity
can be found from  the first  derivative of the
analytical solution as represented by
Eq. (\ref{radiusemdent}),
\begin{equation}
v(t;r_0,v_0,t_0,b) = \frac {d}{dt}r(t;r_0,v_0,t_0,b)
\quad  .
\label{velocityemden}
\end {equation}
The  previous  differential equation (\ref{eqndiff})
can be organized as
\begin{equation}
\frac {d}{dt} r ( t )  = f (r;r_0,v_0,t_0,b)
\quad ,
\label{eqnf}
\end{equation}
and we seek  a power series solution of the form
\begin{equation}
r(t) = a_0 +a_1  (t-t_0) +a_2  (t-t_0)^2+a_3  (t-t_0)^3 + \dots
\quad ,
\label{rtseries}
\end{equation}
see  \cite{Tenenbaum1963,Ince2012}.
The Taylor expansion of Eq. (\ref{eqnf}) gives
\begin{eqnarray}
 f (r;r_0,v_0,t_0,b)=
\nonumber \\
b_0 +b_1  (t-t_0) +b_2  (t-t_0)^2+b_3  (t-t_0)^3 + \dots
\quad ,
\nonumber
\end{eqnarray}
where the values of $b_n$ are
\begin{eqnarray}
b_0=& f (r_0;r_0,v_0,t_0,b)                                    \nonumber \\
b_1=& \frac {\partial} {\partial t} f (r_0;r_0,v_0,t_0,b)      \nonumber \\
b_2=& \frac{1}{2!}\frac {\partial^2} {\partial t^2} f (r_0;r_0,v_0,t_0,b)  \\
b_3=& \frac{1}{3!}\frac {\partial^3} {\partial t^3} f (r_0;r_0,v_0,t_0,b)  \nonumber \\
\ldots &\dots \dots  \nonumber
\end{eqnarray}
The relation between the coefficients $a_n$ and $b_n$ is
\begin{eqnarray}
a_1=& b_0   \nonumber \\
a_2=& \frac{b_1}{2} \nonumber  \\
a_3=& \frac{b_2}{3}  \nonumber \\
\ldots &\dots \dots  \nonumber
\end{eqnarray}
The higher-order  derivatives plus the initial conditions give
\begin{eqnarray}
a_0=& r_0    \nonumber \\
a_1=& v_0   \nonumber \\
a_2=& -\frac{ 9\,{{\it v_0}}^{2}{b}^{2}} { 2\, ( 3\,{b}^{2}+{{\it r_0}}^{2} ) {\it r_0}}\nonumber  \\
a_3=&\frac{ 9\,{{\it v_0}}^{3}{b}^{2} ( 7\,{b}^{2}+{{\it r_0}}^{2} )  }{2\,{{\it r_0}}^{2} ( 3\,{b}^{2}+{{\it r_0}}^{2} ) ^{2} }  \\
\ldots &\dots \dots  \nonumber
\end{eqnarray}
These  are  the coefficient of the {\it second} solution,
which is a power series.

A {\it third } solution can be represented
by a difference equation
which has the following type of recurrence relation
\begin{eqnarray}
r_{n+1} =&  r_n + v_n \Delta t    \nonumber  \\
v_{n+1} =& \frac {{r_{{n}}}^{3}v_{{n}} ( 3\,{b}^{2}+{r_{{n+1}}}^{2} ) ^{3/2}} {( 3\,{b}^{2}+{r_{{n}}}^{2} ) ^{3/2}{r_{{n+1}}}^{3}}
\quad  ,
\label{recursive}
\end{eqnarray}
where  $r_n$, $v_n$, and $\Delta t$ are the temporary
radius,
the velocity, and the interval of time.

The physical units have not yet been specified:
pc for length  and  yr for time
are the units most commonly used by astronomers.
With these units, the initial velocity $v_{{0}}$
is  expressed in $\mathrm{pc \,yr^{-1}}$,
1 yr = 365.25 days,
and should be converted
into   km s$^{-1}$; this means
that   $v_{{0}} =1.02\times10^{-6} v_{{1}}$
where  $v_{{1}}$ is the initial
velocity expressed in
km s$^{-1}$.
In these units, the speed of light is
$c=0.306$  \ pc \ yr$^{-1}$.
The previous analysis covers the case of a symmetrical expansion.
In the framework of the spherical coordinates ($r,\theta,\varphi$)
where $r$  is the radius
($r=0$ refers to the center of the expansion),
$\theta$  is the polar angle with range     $[0-\pi]$, and
$\varphi$ is the azimuthal angle with range $[0-2\pi]$, the
symmetrical  expansion is characterized by the independence
of the advancing radius from $\theta$ and $\varphi$.
We now cover the case of an asymmetrical expansion
in which
the center of the
explosion is at $r=0$ in spherical coordinates  but
does not coincide with the center of the polytrope,
which  is at the distance $z_{asymm}$.
The density
profile of the polytrope, which has now a shifted  origin,
depends now on
the direction of the radius vector.
So, the supernova's shell would evolve in
non-spherical forms.
We align the polar axis  with the line $0-z_{asymm}$.
The symmetry is now around the $z$-axis and the expansion
will be independent of the azimuthal angle
$\varphi$.
The advancing radius will conversely depend on the polar angle
$\theta$.
The  case  of an expansion  that starts  from a given
distance, $z_{asymm}$, from the center of the polytrope
cannot be modeled  by the differential equation (\ref{eqndiff}),
which
is derived  for  a  symmetrical expansion.
It is not possible to find  $R$   analytically  and
a numerical method   should be implemented.
The advancing expansion is computed in a 3D Cartesian
coordinate system ($x,y,z$)  with the center
of the explosion at  (0,0,0).
The degree of asymmetry is evaluated introducing
$R_{eq}$, $R_{up}$  and $R_{down}$
which are the momentari radii in the equatorial plane,
in the polar direction up and in the polar direction
down.
The asymmetry in percentual
is defined by the two ratios
\begin{equation}
a_{up} = \mid \frac{ R_{up} - R_{eq}}{ R_{eq}} \mid *100
\label{asymmetryup}
\quad ,
\end{equation}
and
\begin{equation}
a_{down} = \mid \frac{ R_{down} - R_{eq}}{ R_{eq}} \mid *100
\label{asymmetrydown}
\quad .
\end{equation}
As a reference the  measured asymmetry of \snr
is under  $2 \%$, see \cite{Marcaide2009}.

\subsection{Motion with a power law profile}

The differential equation which models momentum conservation
in the presence of a power law behavior of the density,
as given by (\ref{piecewise}),
is
\begin{eqnarray}
( -4\,{\frac { ( r ( t )  ) ^{3}\rho_{{c}}
\pi }{d-3} ( {\frac {r_{{0}}}{r ( t ) }} ) ^{d}}
+4\,{\frac {\rho_{{c}}\pi \,{r_{{0}}}^{3}}{d-3}}+4/3\,\rho_{{c}}\pi \,
{r_{{0}}}^{3} ) {\frac {\rm d}{{\rm d}t}}r ( t )
\nonumber \\
-4/3 \,\rho_{{c}}\pi \,{r_{{0}}}^{3}{\it v_0}=0
\quad .
\label{eqndiffpowerlaw}
\end{eqnarray}

A {\it first} solution can be found numerically,
see \cite{Zaninetti2011a} for more details.
A {\it second} solution is a  truncated series
about the ordinary point $t=t_0$
which to fourth order has
coefficients
\begin{eqnarray}
a_0=& r_0    \nonumber \\
a_1=& v_0     \nonumber \\
a_2=&  \frac{-3\,{{\it v0}}^{2}} {2\,{\it r_0}} \nonumber \\
a_3=&   \frac{ ( d+7 ) {{\it v_0}}^{3}  } {2\,{{\it r_0}}^{2} } \quad .
\label{coefficientspower}
\end{eqnarray}
A {\it third} approximate solution  can be
found
assuming that
$3 r_0^d r^{4-d}$
$\gg$
$-(4 r_0^3 d-r_0^3 d^2)r$
\begin{eqnarray}
 r(t) =
 ( {r_{{0}}}^{4-d}-\frac{1}{3}d{r_{{0}}}^{4-d}
( 4-d ) \nonumber \\
 + \frac{1}{3}
 ( 4-d ) v_{{0}}{r_{{0}}}^{3-d} ( 3-d )
 ( t-t_{{0}} )  ) ^{\frac{1}{4-d}}
\quad .
\nonumber
\label{asymptotic}
\end{eqnarray}
This is an important approximate result because,
given  the astronomical relation
$r(t)\propto t^{\alpha}$,
we have  $d=4 -\frac{1}{\alpha}$.

\section{Conservation of the relativistic momentum }

\label{secrelativistic}
The  thin layer approximation assumes
that all the swept  mass
during the travel from the initial time, $t_0$,
to the time $t$, resides
in a thin shell of radius $r(t)$ with velocity $v(t)$.
On assuming   a Lane--Emden dependence ($n=5$),
the total  mass $M(r;b)$ comprised between 0 and  $r$
is given by Eq.~(\ref{integralemassa}).
The relativistic  conservation of momentum,
see \cite{French1968,Zhang1997,Guery2010},
is formulated as
\begin{equation}
M(r_0;b) \gamma_0 \beta_0 = M(r;b) \gamma \beta
\quad ,
\nonumber
\end{equation}
where
\begin{equation}
\gamma_0 = \frac{1} {
\sqrt{1-\beta_0^2}
}
\quad ; \qquad
\gamma = \frac{1} {
\sqrt{1-\beta^2}
}
\quad ,
\nonumber
\end {equation}
and
\begin{equation}
\beta_0 =\frac{v_0}{c}
\quad ; \qquad
\beta =\frac{v}{c}.
\nonumber
\end{equation}
The relativistic conservation of momentum
is easily solved for  $\beta$ as a function
of the radius:
\begin{equation}
\beta = \frac{
\sqrt {A ( 3\,{b}^{2}+{r}^{2} ) } ( 3\,{b}^{2}+{r}^{2}
 ) {r_{{0}}}^{3}\beta_{{0}}
 }
 {A}
\quad ,
\nonumber
\label{eqnbeta}
\end{equation}
with
\begin{eqnarray}
A = -27\,{b}^{6}{r}^{6}{\beta_{{0}}}^{2}+27\,{b}^{6}{\beta_{{0}}}^{2}{r_{{0
}}}^{6}-27\,{b}^{4}{r}^{6}{\beta_{{0}}}^{2}{r_{{0}}}^{2}+27\,{b}^{4}{r
}^{2}{\beta_{{0}}}^{2}{r_{{0}}}^{6}
\nonumber \\
-9\,{b}^{2}{r}^{6}{\beta_{{0}}}^{2}
{r_{{0}}}^{4}+9\,{b}^{2}{r}^{4}{\beta_{{0}}}^{2}{r_{{0}}}^{6}+27\,{b}^
{6}{r}^{6}+27\,{b}^{4}{r}^{6}{r_{{0}}}^{2}+9\,{b}^{2}{r}^{6}{r_{{0}}}^
{4}+{r}^{6}{r_{{0}}}^{6}
\nonumber
\quad .
\end{eqnarray}
Inserting
\begin{equation}
\beta = \frac{1}{c}{\frac {d}{dt}}r( t ),
\nonumber
\end{equation}
the relativistic conservation of momentum
can be written as the differential
equation
\begin{equation}
\frac{
4\,{b}^{3} ( r ( t )  ) ^{3}\rho\,\pi \,\sqrt {3
}{\frac {d}{dt}}r ( t )
}
{
( 3\,{b}^{2}+ ( r ( t )  ) ^{2} ) ^{
3/2}c\sqrt {-{\frac { ( {\frac {d}{dt}}r ( t )
 ) ^{2}}{{c}^{2}}}+1}
}
= \frac {
4\,{b}^{3}{r_{{0}}}^{3}\rho\,\pi \,\sqrt {3}\beta_{{0}}
}
{
( 3\,{b}^{2}+{r_{{0}}}^{2} ) ^{3/2}\sqrt {-{\beta_{{0}}}^{
2}+1}
}
\quad .
\label{eqndiffrel}
\end{equation}
This first order differential equation
can be solved by separating the variables:
\begin{equation}
\int_{r_0}^r
\frac{A}{
\sqrt {A ( 3\,{b}^{2}+{r}^{2} ) } ( 3\,{b}^{2}+{r}^{2}
 ) {r_{{0}}}^{3}\beta_{{0}}
 }
\,dr
= c (t-t_0)
\quad .
\label{nlequation}
\end{equation}
The previous integral does not have an analytical solution
and we treat the previous result as a non-linear
equation to be
solved numerically.
The differential equation has a truncated series solution
about the ordinary point $t=t_0$
which to fifth order is
\begin{equation}
r_s( t )=
\sum _{n=0}^{4}a_{{n}}{(t-t_0)}^{n}
\quad .
\label{rtseriesrel}
\end{equation}
The coefficients are
\begin{eqnarray}
a_0=& r_0    \nonumber \\
a_1=& c\beta_{{0}}   \nonumber \\
a_2=& \frac {9\,{b}^{2}{\beta_{{0}}}^{2}{c}^{2} ( {\beta_{{0}}}^{2}-1 )}{ 2\,r_{{0}} ( 3\,{b}^{2}+{r_{{0}}}^{2} ) }              \\
a_3=&\frac{9\,{c}^{3} ( \beta_{{0}}-1 )  ( \beta_{{0}}+1 )
 ( 12\,{b}^{2}{\beta_{{0}}}^{2}-7\,{b}^{2}-{r_{{0}}}^{2} )
{\beta_{{0}}}^{3}{b}^{2}} {2\, ( 3\,{b}^{2}+{r_{{0}}}^{2} ) ^{2}{r_{{0}}}^{2}} \nonumber \\
a_4=&\frac{9\,{b}^{2} ( \beta_{{0}}-1 )  ( \beta_{{0}}+1 )
B{\beta_{{0}}}^{4}{c}^{4}} {8\,{r_{{0}}}^{3} ( 3\,{b}^{2}+{r_{{0}}}^{2} ) ^{3}} \nonumber  \\
where ~B=& 756\,{b}^{4}{\beta_{{0}}}^{4}-927\,{b}^{4}{\beta_{{0}}}^{2}-117\,{b}^{
2}{\beta_{{0}}}^{2}{r_{{0}}}^{2}+231\,{b}^{4}+69\,{b}^{2}{r_{{0}}}^{2}
+4\,{r_{{0}}}^{4}  \nonumber
\end{eqnarray}
The velocity approximated to the fifth order is
\begin{equation}
v_s( t ) =
\sum _{n=1}^{4}{\frac {a_{{n}}{(t-t_0)}^{n}n}{(t-t_0)}}
\quad .
\label{vseries}
\end{equation}
The presence of an analytical  expression for $\beta$
as given  by Eq. (\ref{eqnbeta})  allows the
recursive solution
\begin{eqnarray}
r_{n+1} =   &  r_n + c \beta_n \Delta t    \nonumber  \\
\beta_{n+1} =&
\frac
{
\sqrt {A ( 3\,{b}^{2}+{r_{{n+1}}}^{2} ) } ( 3\,{b}^{2}
+{r_{{n+1}}}^{2} ) {r_{{n}}}^{3}\beta_{{n}}
}
{
A
}    \\
with ~A=&
27\,{b}^{6}{\beta_{{n}}}^{2}{r_{{n}}}^{6}-27\,{b}^{6}{\beta_{{n}}}^{2}
{r_{{n+1}}}^{6}+27\,{b}^{4}{\beta_{{n}}}^{2}{r_{{n+1}}}^{2}{r_{{n}}}^{
6}
\nonumber    \\
-27\,{b}^{4}{\beta_{{n}}}^{2}{r_{{n+1}}}^{6}{r_{{n}}}^{2}
& +9\,{b}^{2}
{\beta_{{n}}}^{2}{r_{{n+1}}}^{4}{r_{{n}}}^{6}-9\,{b}^{2}{\beta_{{n}}}^
{2}{r_{{n+1}}}^{6}{r_{{n}}}^{4}+27\,{b}^{6}{r_{{n+1}}}^{6}
\nonumber  \\
+27\,{b}^{4}
{r_{{n+1}}}^{6}{r_{{n}}}^{2}
& +9\,{b}^{2}{r_{{n+1}}}^{6}{r_{{n}}}^{4}+{r
_{{n+1}}}^{6}{r_{{n}}}^{6}
\nonumber
\quad  ,
\label{recursiverel}
\end{eqnarray}
where  $r_n$, $\beta_n$ and $\Delta t$
are the temporary  radius,
the relativistic $\beta$
factor, and the interval of time, respectively.
Up to now we have taken the time interval $t-t_0$ to be that as seen
by an observer on earth.
For an   observer which moves on the expanding shell,
the proper time $\tau^*$ is
\begin{equation}
\tau^* = \int_{t_0}^{t} \frac {dt}{\gamma} =  \int_{t_0}^{t}
\sqrt{1-\beta^2} dt
\quad ,
\nonumber
\end{equation}
see  \cite{Larmor1897,Lorentz1904,Einstein1905,Macrossan1986}.
In  the series solution framework,
$\beta=\frac{v_s}{c}$,
and   $v_s$ is given by Eq. (\ref{vseries}).
A measure  of the time dilation
is  given  by
\begin{equation}
D = \frac{\tau^*}{t-t_0}
\quad ,
\nonumber
\end {equation}
with   $0 < D < 1 $.
It is interesting to point out  that the time dilation
here analyzed  is not connected
with the ``cosmological time dilation in GRB''
which, conversely, is related to the cosmological
redshift, see \cite{Kocevski2013,Zhang2013}
for two diametrically opposed  points of view.
An application of  the time dilation
is the  decay of a radioactive  isotope as  modeled by the
following law for remnant particles in
the laboratory framework
\begin{equation}
N(t) = N_0 e^{-\frac{(t-t_0)}{\tau}}
\nonumber
\label{ntradioactivelab}
\quad ,
\end{equation}
where
$\tau$ is the proper lifetime,
$N_0$  is the number of nuclei at  $t=t_0$
and the  half life is  $T_{1/2}= ln(2) \; \tau$.
In a frame that is moving with the shell,
the decay law is
\begin{equation}
N(t) = N_0 e^{-\frac{\tau^*}{\tau}}
\nonumber
\label{ntradioactivesn}
\quad .
\end{equation}
This theory is used to explain the lifetime of the
muons in
cosmic rays:
in that case, $\gamma \approx 8.4 $ and
$\tau = 2.196\times10^{-6}$\ s,
see \cite{Rossi1941,Frisch1963,PDG2012}.
In SR, the total energy of a particle  is
\begin{equation}
E = m c^2 = m_0 \gamma c^2
\quad ,
\nonumber
\end{equation}
where $m_0$ is the rest mass.
The relativistic kinetic energy is
\begin{equation}
KE = m_0 c^2 (\gamma-1)
\quad ,
\nonumber
\end{equation}
where the rest energy has been subtracted
from the total energy.
In order to have  a simple expression for the
velocity as a function of time,
we deduce  a series expansion for the radius
as a function of time   limited  to the third order.
The relativistic kinetic energy is therefore
\begin{equation}
KE = m_0 c^2
(
{\frac {1}{\sqrt {1 - ( c\beta_{{0}}+9\,{\frac {{b}^{2}{\beta_{{0}}
}^{2}{c}^{2} ( {\beta_{{0}}}^{2}-1 )  ( t-t_{{0}}
 ) }{r_{{0}} ( 3\,{b}^{2}+{r_{{0}}}^{2} ) }} )
^{2}{c}^{-2}}}}
-1)
\quad .
\nonumber
\end{equation}

\section{Classical astrophysical applications}

\label {applicationsclassic}
This section introduces:
the SN chosen for testing purposes,
the astrophysical environment connected with the selected SN,
two types of fit
commonly used to model the radius--time relation
in SN,
the
application of the results obtained for
the Lane--Emden density  profile  to the selected  SN,
the asymmetric explosion, and the case of CSM characterized
by a power law.

\label{secobservations}

\subsection{The data}

The data of \snr,
radius in  pc and elapsed time  in years,
can be found in  Table 1 of  \cite{Marcaide2009}.
The instantaneous velocity of expansion
can be deduced from the formula
\begin{equation}
v_i = \frac
{r_{i+1} - r_i}
{t_{i+1} - t_i}
\quad ,
\nonumber
\label{vdiscrete}
\end{equation}
where $r_i$ is the radius and $t_i$  is
the time
at the position $i$.
The uncertainty in the instantaneous   velocity
is found by implementing the error
propagation equation, see \cite{Bevington2003}.
A discussion of the
thickness of the radio shell
in \snr in the framework of
a reverse shock \cite{Chevalier1982a,Chevalier1982b}
can be found in \cite{Bartel2007}.
The thickness of the radio shell can also be explained
in the framework of the image theory, see
Section 6.3 in \cite{Zaninetti2011a}.

\subsection{Astrophysical Scenario}

The progenitor of \snr  was a K-supergiant star, see \cite{Aldering1994}
 and probably formed a binary system with an B-supergiant companion
star, see \cite{Maund2004}.
These massive stars have strong
stellar winds, and blow huge bubbles (of $\approx$  20 to 40 pc in size) in their
lives.
From an analytic approximation \cite{Weaver1977}
obtained a formula for the radius of the bubble, their eqn.(21),
see too their Figure 3.
The inner of the bubble has very low density, and the border of the bubble
is the wall of a relatively dense shell which is in contact with the ISM.
The circumstellar envelope of the Pre \snr with which is interacting the
SN shock front is a small structure within the big bubble
created by the strong
stellar winds of the SN progenitor
(and probably of its binary companion) during
its life. Therefore this envelope of the Pre \snr
would be the product of a
recent event of stellar mass ejection suffered by the Pre \snr.
That is to
say that the SN shock wave interacts with a CSM created by
Pre supernova mass loss. In this respect,
\cite{Schmidt1993}
gave evidences that significant mass loss
 had taken place before the explosion, see also \cite{Smith2008}.
 In the scenario that the Pre \snr
formed an interacting binary system,
this can be interpreted in terms of a
process of mass transfer.
It is possible that this type of supernova originates
in interacting binary systems.

\subsection{Two types of fit}

The quality of the fits is measured by the
merit function
$\chi^2$
\begin{equation}
\chi^2  =
\sum_j \frac {(r_{th} -r_{obs})^2}
             {\sigma_{obs}^2}
\quad ,
\nonumber
\label{chisquare}
\end{equation}
where  $r_{th}$, $r_{obs}$ and $\sigma_{obs}$
are the theoretical radius, the observed radius, and
the observed uncertainty, respectively.
A {\em first} fit
can be done by assuming  a  power law
dependence  of the type
\begin{equation}
r(t) = r_p t^{\alpha_p}
\nonumber
\label{rpower}
\quad ,
\end{equation}
where the  two parameters $r_p$ and  $\alpha_p$
as well their  uncertainties
can be found
using the recipes  suggested in
\cite{Zaninetti2011a}.
A {\em second}  fit   can be done by assuming
a piecewise function as  in
Fig.~4 of \cite{Marcaide2009}
\begin{equation}
 r(t)   = \{ \begin{array}{ll}
             r_{br}(\frac{t}{t_{br}})^{\alpha_1} &
              \mbox {if $t \leq t_{br} $ } \nonumber \\
             r_{br}(\frac{t}{t_{br}})^{\alpha_2} &
               \mbox {if $t > t_{br} $. } \nonumber
            \end{array}
            .
\label{piecewisefit}
\end{equation}
This type of fit requires the determination
of four parameters: $t_{br}$ the break time,
$r_{br}$ the radius of expansion at
$t=t_{br}$,
and the exponents $\alpha_1$ and $\alpha_2$ of the two phases.
 The parameters of these two
 fits  as well the $\chi^2$  can  be found in
 Table~\ref{datafit}.
\begin{table}
\caption
{
Numerical values of the parameters
of the fits and
$\chi^2$;
$N$  represents the number of
free parameters.
}
 \label{datafit}
 \[
 \begin{array}{cccc}
 \hline
 \hline
 \noalign{\smallskip}
  N
& values  & \chi^2        \\
 \noalign{\smallskip}
 \hline
 \noalign{\smallskip}
  &power~law~as~a~fit   &   &\\ \noalign{\smallskip}
  2   & \alpha_p = 0.82 \pm 0.0048  & 6364  \\ \noalign{\smallskip}
 ~    &r_p = (0.015 \pm 0.00011) ~{\mathrm{pc}}  &   \\
 \noalign{\smallskip}
 \hline
        & piecewise~fit   &   &\\ \noalign{\smallskip}
 4   & \alpha_1 = 0.83 \pm 0.01  & 32

 \\
   & \alpha_2 = 0.78 \pm 0.0077; & ~
 \\

 \noalign{\smallskip}
 ~   & r_{br} = 0.05~{\mathrm{pc}};
 t_{br}=4.10~{yr}
&
 \\
 \noalign{\smallskip}
 \hline
  &Plummer~profile~,\eta=6  &   &\\ \noalign{\smallskip}
  2   & b=0.0045~{\mathrm{pc}} ; r_{0} = 0.008~{\mathrm{pc}};v_0=19500\frac{km}{s} &  265
\\ \noalign{\smallskip}
 \hline
  &Lane--Emden ~profile  &   &\\ \noalign{\smallskip}
  2   & b=0.00367 ~{\mathrm{pc}}; r_{0} = 0.008~{\mathrm{pc}};v_0=19500\frac{km}{s} &  471
\\
\noalign{\smallskip} \hline
  &Power~law~profile  &   &\\ \noalign{\smallskip}
  2   & d=2.93; r_{0} = 0.0022~{\mathrm{pc}}; &  276
\\
\noalign{\smallskip} ~   & t_{0}=0.249~{yr}; v_0=100000\frac{km}{s} & ~
\\
\noalign{\smallskip} \hline\hline
 \end{array}
 \]
 \end {table}

\subsection{The Lane--Emden  case}

\begin{figure*}
\begin{center}
\includegraphics[width=7cm ]{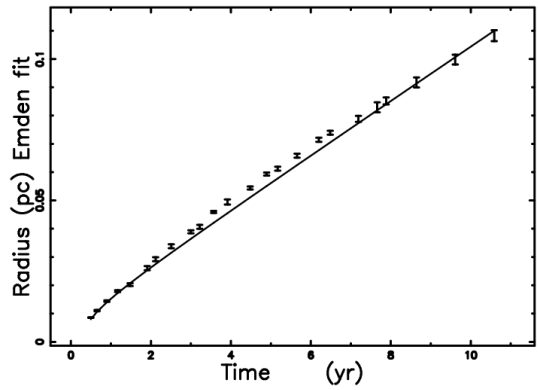}
\end {center}
\caption
{
Theoretical radius as given
by Eq.~(\ref{radiusemdent})
(full line),
with data as in Table~\ref{datafit}.
The  astronomical data of \snr
are represented with vertical error bars.
}
\label{1993pc_fit_emden}
    \end{figure*}

The radius  of \snr
which represents the  momentum conservation
in a Lane--Emden profile  of density
is  reported  in Fig.~\ref{1993pc_fit_emden};
$r_0$ and $t_0$ are  fixed by the observations
and the two free  parameters are $b$ and $v_0$.

Fig.~\ref{1993pc_fit_series} compares
the theoretical solution and the series expansion
about the  ordinary point $t_0$.
The range of time in which  the series solution
approximates the analytical solution is limited.
Fig.~\ref{1993pc_fit_recurs} compares
the theoretical solution and
the recursive solution
as represented by Eq. \ref{recursive}.
The recursive solution
approximates the analytical solution
over all the range of time
considered and the error at $t=10$\ yr is $\approx$ 0.6$\%$ when
$\Delta t=0.05$ yr and $\approx$ 0.1$\%$ when
$\Delta t=0.0083$ yr.
The time evolution of the velocity is reported
in Fig.~\ref{1993pc_velocity_emden}.
The asymmetrical case in which there is a distance, $z_{asymm}$,
from the center of the
polytrope  and the center of the expansion
is clearly outlined in Fig.
\ref{1993secasymm}. In this figure we have two  sections
of the expansion in the plane
connecting $z_{asymm}$ with the center of the expansion
with  $z_{asymm}$ increasing from 0 (the  circular section)
to 0.00367 pc (the asymmetrical section).

Fig.~\ref{sn1993j3d} reports the complex structure
of the 3D
advancing surface.
The point of view
of the observer
is  parametrized by
the Euler angles $(\Phi, \Theta, \Psi)$.
\subsection{Plummer and power law cases}

The numerical solution of the differential equation
connected with the Plummer-like profile, $\eta=6$,
is reported in  Fig.~\ref{snr_1993_plummer} when
the data of Table \ref{datafit} are adopted.

A comparison with the power law behavior for the CSM
is reported in Fig.~\ref{snr_1993_emden_power}
which is built from the data in Table \ref{datafit}.
The series solution for the power law dependence of the CSM
with coefficients as given by  Eq. (\ref{coefficientspower})
is not reported  because
the  range in time
of reliability is limited to $t-t_0 \approx 0.0003$\ yr.

\section{Relativistic astrophysical applications}
\label{applications}
We now apply the relativistic solutions  derived so far to \snr.
The initial observed velocity, $v_0$,  as deduced
from  radio observations,
see \cite{Marcaide2009}, is $v_0 \approx 20000\,\velu$  at
$t_0 \approx 0.5 \, \mathrm{ yr}$.
We now reduce the initial time $t_0$ and we increase the velocity
up to the relativistic regime, $t_0=10^{-4}\, \mathrm{yr}$,
and $v_0= 100000\,\velu$.
This choice of parameters allows fitting
the observed radius--time
relation that  should be reproduced.
The  data used in the simulation are
shown in Table \ref{datafitrel}.
\begin{table}
\caption
{
Numerical values of the parameters
used in  three relativistic solutions.
}
 \label{datafitrel}
 \[
 \begin{array}{c}
 \hline
 \hline
 \noalign{\smallskip}
 parameters      \\
  t_0=10^{-4}~\mathrm{yr}~;
  r_0=0.0033~\mathrm{pc} ~;
~\beta_0=0.3333 ~;
~ b=0.004~ \mathrm{pc}
\\
\noalign{\smallskip} \hline
 \end{array}
 \]
 \end {table}
The relativistic numerical solution of Eq. (\ref{nlequation})
is reported in Fig.~\ref{sol_analytical},
the relativistic series solution as given
by (\ref{rtseriesrel}) is reported in Fig.~\ref{sol_series},
and Fig.~\ref{sol_recursive} contains the
recursive solution  as given by Eq. (\ref{recursiverel}).

Fig.~\ref{timecontract} reports
a 2D map of the parameter $D$ which
parametrizes the time dilation.

Fig.~\ref{decays} reports the temporal evolution
in the number of $^{56}$Ni
($\tau $ = 8.757\ d)
in the laboratory frame at rest and in
the frame which is moving with the SN.
Fig.~\ref{energy_time} reports
the temporal evolution of the  relativistic  kinetic energy
for a proton.

\section{Conclusions}

\begin{figure*}
\begin{center}
\includegraphics[width=7cm ]{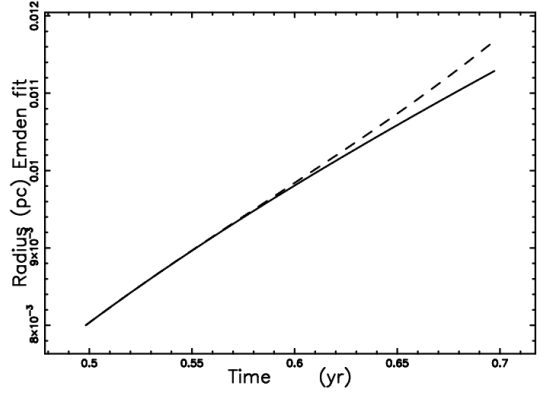}
\end {center}
\caption
{
Theoretical radius as given
by Eq. (\ref{radiusemdent}) (full line)
and series solution as
given by Eq. (\ref{rtseries}) (dashed line).
Data as in Table \ref{datafit}.
}
\label{1993pc_fit_series}
    \end{figure*}

\begin{figure*}
\begin{center}
\includegraphics[width=7cm ]{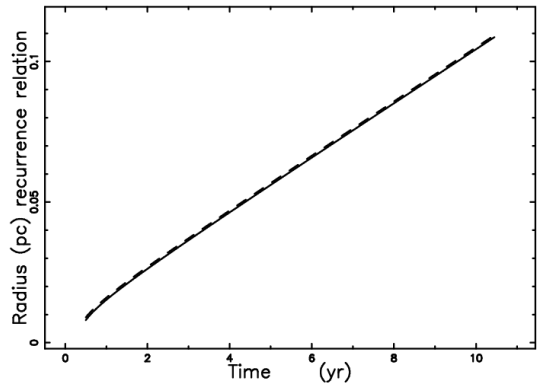}
\end {center}
\caption
{
Theoretical radius as given
by Eq. (\ref{radiusemdent}) (full line)
and recursive solution as
given by Eq. (\ref{recursive})
when $\Delta t=0.05 \mathrm{yr}$ (dashed line).
Data as in Table \ref{datafit}.
}
\label{1993pc_fit_recurs}
    \end{figure*}

\begin{figure*}
\begin{center}
\includegraphics[width=7cm ]{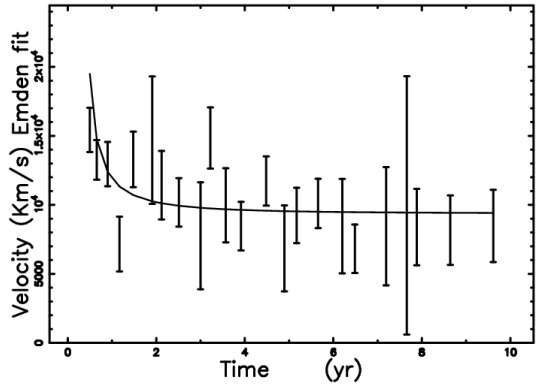}
\end {center}
\caption
{
Instantaneous velocity
of \snr with
uncertainty  and    theoretical  velocity
as given by  Eq. (\ref{velocityemden}) (full line).
Data as in Table \ref{datafit}.
}
\label{1993pc_velocity_emden}
    \end{figure*}

\begin{figure}
\begin{center}
\includegraphics[width=6cm]{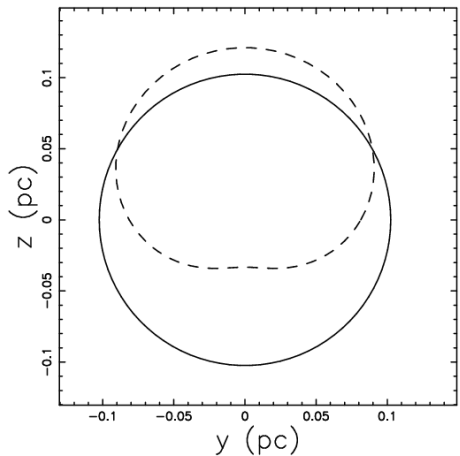}
\end{center}
\caption{
Sections  of asymmetrical SN with the same initial parameters of \snr
in the plane which contains the center of the polytrope
when  $z_{asymm}=0$\ pc (full line) and
$z_{asymm}=-0.00367$\ pc
(dashed  line).
Data as in Table \ref{datafit}.
In the asymmetrical case (dashed line) the degrees
of asymmetry are  $a_{up}= 32\%$
and $a_{down}= 32\%$,
see formulae (\ref{asymmetryup}) and
(\ref{asymmetrydown}).
}

\label{1993secasymm}%
    \end{figure}

\begin{figure}
\begin{center}
\includegraphics[width=6cm]{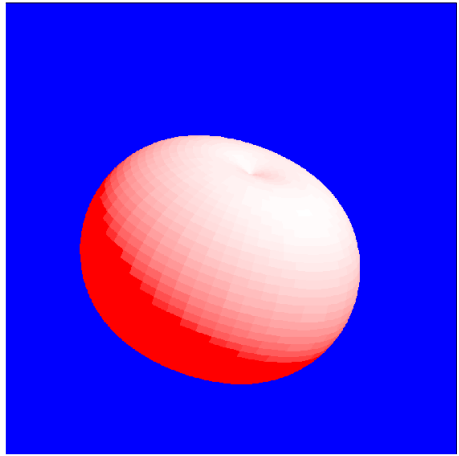}
\end{center}
\caption
{
The 3D advancing surface
of an asymmetrical SN with the same initial parameters of \snr.
The three Eulerian angles characterizing the
      point of view are
     $ \Phi   =70 ^{\circ }$,
     $ \Theta =70 ^{\circ }$
and  $ \Psi   =70 ^{\circ }$
}
\label{sn1993j3d}%
    \end{figure}

\begin{figure*}
\begin{center}
\includegraphics[width=7cm ]{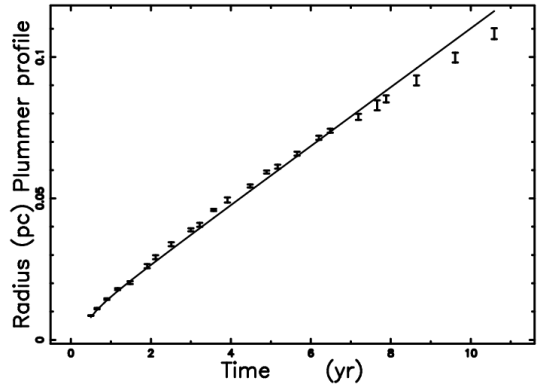}
\end {center}
\caption
{
Theoretical radius for the Plummer-type profile as obtained
by the  solution of  the nonlinear
equation connected with (\ref{eqndiffplummer})
(full line).
Data as in Table \ref{datafit}.
}
\label{snr_1993_plummer}
    \end{figure*}

\begin{figure*}
\begin{center}
\includegraphics[width=7cm ]{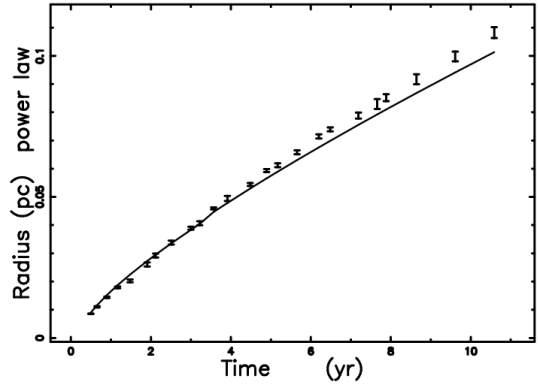}
\end {center}
\caption
{
Theoretical radius for the power law case as obtained
by the  solution of  the nonlinear
equation connected with (\ref{eqndiffpowerlaw})
(full line).
Data as in Table \ref{datafit}.
The astronomical data of \snr are represented with
vertical error bars.
}
\label{snr_1993_emden_power}
    \end{figure*}

\begin{figure*}
\begin{center}
\includegraphics[width=7cm ]{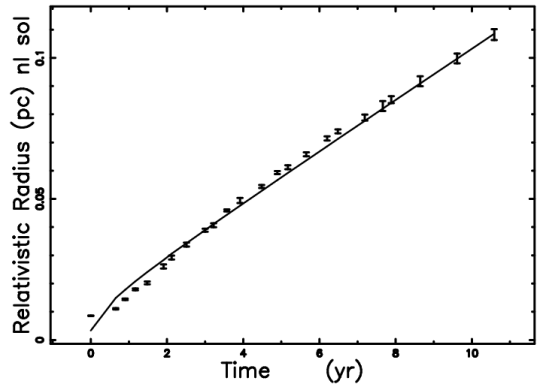}
\end {center}
\caption
{
Theoretical relativistic radius as solution of the
non-linear equation (\ref{nlequation}) (full line),
with data as in Table~\ref{datafitrel}.
\label{sol_analytical}
}
    \end{figure*}

\begin{figure*}
\begin{center}
\includegraphics[width=7cm ]{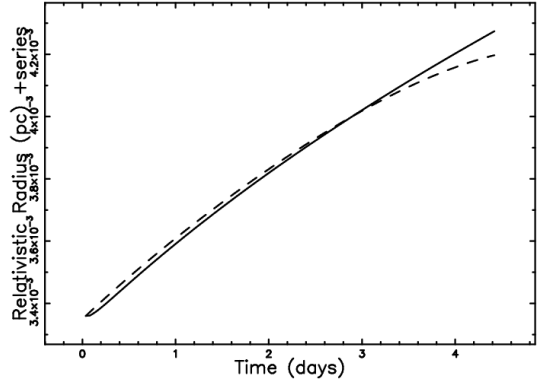}
\end {center}
\caption
{
Theoretical relativistic radius as solution of the
non-linear equation (\ref{nlequation}) (full line),
and series solution as
given by Eq. (\ref{rtseriesrel}) (dashed line).
Data as in Table~\ref{datafitrel}.
The time is expressed in days.
}
\label{sol_series}
    \end{figure*}

\begin{figure*}
\begin{center}
\includegraphics[width=7cm ]{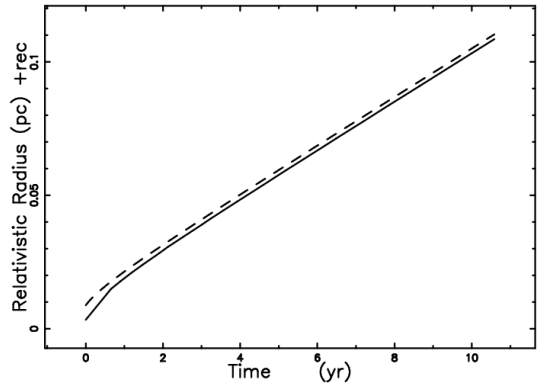}
\end {center}
\caption
{
Theoretical relativistic radius as solution of the
non-linear equation (\ref{nlequation}) (full line),
and recursive solution as
given by Eq. (\ref{recursiverel})
when $\Delta t=0.053 \mathrm{yr}$ (dashed line).
Data as in Table~\ref{datafitrel}.
}
\label{sol_recursive}
    \end{figure*}

{\bf Classic Case:}
The thin layer approximation  which models
the expansion  in an self-gravitating medium
of the Lane--Emden type ($n=5$)
can be modeled
by a differential  equation of the first order
for the radius as a function of  time.
This  differential  equation has
an analytical solution represented by Eq.
(\ref{radiusemdent}).
A power law series, see Eq. (\ref{rtseries}),
can model the solution
of the Lane--Emden type
for a limited range of time,
see Fig.~\ref{1993pc_fit_series}.
Conversely, a  recursive solution for the first
order differential equation,
as represented by Eq. (\ref{recursive}), approximates
quite well the analytical solution of the Lane--Emden type
 and at the time
of $t=10$ yr a precision of four digits is reached
when $\Delta t=10^{-3}$ yr.
The goodness of the results as given by the solutions
of the three differential equations is evaluated in
Table~\ref{datafit}.
The smallest $\chi^2$ is obtained by the Plummer-like ($\eta=6$)
profile,
followed by the power law profile and the Lane--Emden type ($n=5$)
profile.
The two-piece fit has the smallest $\chi^2$
but requires four parameters and does not have a physical basis.
The disadvantages  of the power law dependence in the CSM
are:
(i) there is a two-piece dependence
at $r=r_0$ which was introduced in order to avoid a pole,
(ii) there is no analytical solution,
(iii) the series solution has a narrow range of reliability.

{\bf Auto-gravitating medium:}
The previous analysis raises a question:
is the CSM around an SN really self-gravitating?
In order to answer this question, the CSM should be carefully
analyzed by astronomers in order to detect the presence
of gradients in density or pressure or temperature.
In the case of  a Lane--Emden ($n=5$) CSM,
the density of the medium around \snr , see
Eq. (\ref{densita5}),
decreases
by a factor $\approx$ 21 going from $r_0=0.008$~pc
to $r=0.1$~pc.
Over the same distance, the pressure,
 see Eq. (\ref{pressure}),
decreases by a factor $\approx$ 40
and the temperature,
see Eq. (\ref{temperature}),
by a factor $\approx$ 1.8.
The  presence or absence of
a magnetic field should
be also confirmed.

{\bf Nature of the CSM:}
The obtained total mass in the three models can be
interpreted as the stellar mass ejected by
the pre-SN right before the explosion
or the stellar mass involved with
the interaction of the binaries,
see Table \ref{massfwhm}.
The size of the
pre-\snr  envelope, i.e. the FWHM,
is an important data that could be
related with the size of the Roche lobes of the hypothetical binary system,
see Table \ref{massfwhm}.
\begin{table}
\caption
{
Numerical values
of the swept total mass when $n_0= 10^7 \mathrm{cm}^{-3} $  and the FWHM.
The parameter $n_7$ represents the number density expressed
in $10^7 \mathrm{cm}^{-3}$ units.
}
 \label{massfwhm}
 \[
 \begin{array}{ccc}
 \hline
 \hline
 \noalign{\smallskip}
 Model
& Total ~Mass
& FWHM        \\
 \noalign{\smallskip}
 \hline
 \noalign{\smallskip}
 Plummer~profile~,\eta=6
 &  M(0.1;0.0045)=0.402\, n_7\, M_{\sun}
 & 0.0079~{\mathrm{pc}}
\\
\noalign{\smallskip}
  Emden~profile
 &  M(0.008;0.00367)=0.178 \,n_7 M_{\sun}
 & 0.0071~{\mathrm{pc}}
 \\
 \noalign{\smallskip}
  Emden~profile
 &  M(0.1;0.00367) =0.368 \,n_7 M_{\sun}
 & 0.0071~{\mathrm{pc}}
 \\
 \noalign{\smallskip}
 Power~law~profile
 &  M(0.1;0.0022,2.93)=0.217  \, n_7 M_{\sun}
 & 0.0092~{\mathrm{pc}}
 \\
 \noalign{\smallskip}
\hline\hline
 \end{array}
 \]
 \end {table}
Here we have chosen, in the three models,
a central density of
$n_0= 10^7 \mathrm{cm}^{-3} $.
As a comparison \cite{Suzuki1995}, in order to model
the X-ray observations, quotes a central density
of $5.7 \,10^{-4} g \mathrm{cm}^{-3}$ at a distance
of $\approx 10^{-5} \mathrm{pc}$ which means
$n_0= 2.45\, 10^{10}   \mathrm{cm}^{-3} $.
An astrophysical evaluation of  $n_7$
can be done by imposing that the swept mass
is a fraction $f$, $0 <  f < 1$,  of the mass of the progenitor
which is $5\,M_{\sun}$, see
\cite{Woosley1994};
in the case of the Emden profile
$n_7=13.57 \, f$.
With this evaluation  the
velocity dispersions to maintain the
dynamic equilibrium, as given by formula (\ref{vvirial}),
are of order
$3.161\,\sqrt {f} \frac{km}{s}$.
The previous table allows a fast evaluation of the
final velocity, $v$, in the framework of the momentum conservation
$v=v_0 \frac{M(0.008;0.00367)}{M(0.1;0.00367)}= 15437 \frac{0.1778}{0.3686}
\frac{km}{s}= 7447 \frac{km}{s}$.

{\bf Asymmetry:}
In the binary system scenario, both stars had probably a
common envelope centered at the gravity center of the system.
Hence, the SN exploded at a distance $\approx$ a/2
(assuming two stars of similar masses) with respect to
the center of the density distribution of the CSM, where
a is the distance between the stars of the binary system.
In the case of \snr  the expansion
shows a  circular expansion
over 10 years and the asymmetry is under  $2 \%$.
This means,  in the binary system scenario,
that the distance, d,  between initial point of the expansion and
the center of the Lane--Emden ($n=5$) CSM  should be
$ d \leq 0.037 \, b$ or $ d \leq  0.00013 pc $.
The distance a/2  is therefore equalized to our parameter
d  which means  $ a \leq  0.00026 pc $.

\begin{figure*}
\begin{center}
\includegraphics[width=7cm ]{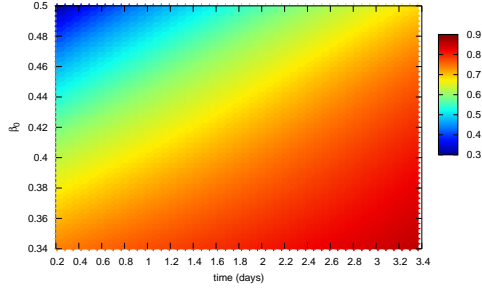}
\end {center}
\caption
{
Map of time dilation  as represented by $D$
as a function of time (in days) and $\beta_0$.
Data as in Table~\ref{datafitrel}.
}
\label{timecontract}
    \end{figure*}

\begin{figure*}
\begin{center}
\includegraphics[width=7cm ]{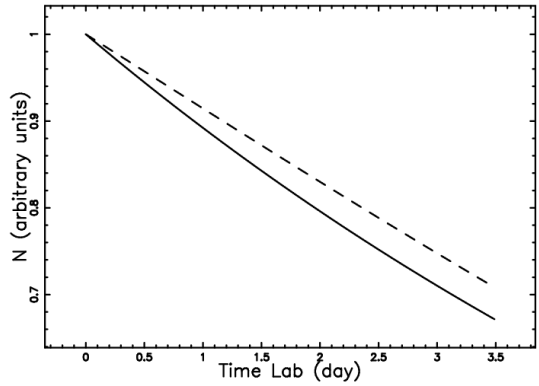}
\end {center}
\caption
{
Number of  nuclei
of $^{56}$Ni in the  inertial frame of the laboratory
(full line) and in the frame that is moving with the SN
(dashed line).
Data as in Table~\ref{datafitrel}.
}
\label{decays}
    \end{figure*}

\begin{figure*}
\begin{center}
\includegraphics[width=7cm ]{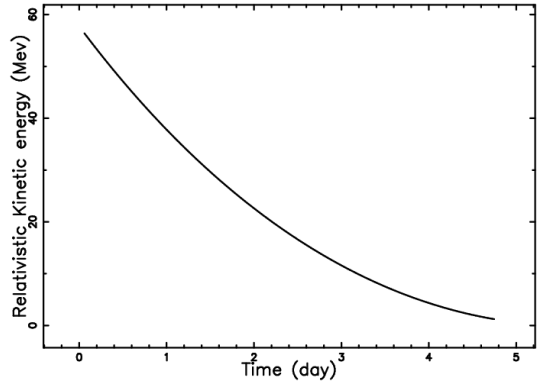}
\end {center}
\caption
{
Relativistic kinetic energy of a proton in Mev
as a function of time (in days).
Data as in Table~\ref{datafitrel}.
}
\label{energy_time}
    \end{figure*}

{\bf Relativistic case:}
The temporal  evolution of a SN in an self-gravitating medium
of the Lane--Emden type can be found by applying the
conservation of relativistic momentum in the thin layer
approximation.
This  relativistic invariant is  evaluated as
a differential equation of the first order, see
Eq. (\ref{eqndiffrel}).
Three different relativistic solutions   for the radius as a function
of time  are derived:
(i) a numerical  solution,
see Eq.~(\ref{nlequation})
which covers the range $10^{-4}\mathrm{ yr} < \, t < 10\, \mathrm{yr} $
and
fits the observed radius--time relation
for \snr;
(ii) a series solution, see Eq.~(\ref{nlequation}),
which has a limited
range of validity,
$10^{-4}\mathrm{ yr} < \, t < 1.2\times10^{-2} \, \mathrm{yr} $;
(iii) a recursive solution,
see Eq.~(\ref{recursiverel}),
in which the  desired accuracy  is  reached
by  decreasing the time step $\Delta t$.
The relativistic results here presented
model \snr and   are obtained with
an  initial velocity of $v_0=100 000\velu$ or $\beta_0=0.333$
or $\gamma=1.06$.
The time dilation is evaluated and then
applied to the decay of  $^{56}$Ni,
see Fig.~\ref {timecontract}.
The relativistic kinetic energy of a proton
is computed
and the  temporal evolution  in Mev outlined,
see Fig.~\ref{energy_time}.
\newpage

\newpage

%
%

%

%
%
%
%

\end{document}